\documentclass[prd,twocolumn,showpacs,preprintnumbers,floatfix,amsmath,amssymb,amsfonts]{revtex4}

\usepackage{graphicx}
\usepackage{dcolumn}
\usepackage{epsfig}
\usepackage{relsize}
\RequirePackage{xspace} 

\newcommand{\BABARPubYear}    {08}
\newcommand{\BABARPubNumber}  {048}   
\newcommand{\SLACPubNumber}   {13508}

\def\babar   {\mbox{\slshape B\kern-0.1em{\smaller A}\kern-0.1em B\kern-0.1em{\smaller A\kern-0.2em R}}}
\def\peptwo  {PEP-II}
\def\epem    {\ensuremath{e^+e^-}\xspace}
\def\invfb   {\ensuremath{\mbox{\,fb}^{-1}}\xspace}
\def\qqbar   {\ensuremath{q\overline q}\xspace}
\def\de      {\mbox{$\Delta E$}\xspace}
\def\mes     {\mbox{$m_{\rm ES}$}\xspace}
\def\Y       {\ensuremath{\Upsilon{\rm(4S)}}\xspace}
\def\B       {\ensuremath{B}\xspace}
\def\Bbar    {\ensuremath{\kern 0.18em\overline{\kern -0.18em B}}\xspace}
\def\BB      {\ensuremath{B\Bbar}\xspace}
\def\Bub     {\ensuremath{B^-}\xspace}
\def\Bm      {\ensuremath{\Bub}\xspace}
\def\BpBm    {\ensuremath{B^+B^-}\xspace}
\def\Bzb     {\ensuremath{{\Bbar}\kern0.03em {\raise.3ex\hbox{$^0$}}}\xspace}
\def\BoBob   {\ensuremath{B^0 {\kern -0.16em \Bzb}}\xspace}

\def\DJ      {\ensuremath{D_J}\xspace}
\def\DJz     {\ensuremath{D_J^0}\xspace}
\def\Dp      {\ensuremath{D^+}\xspace}
\def\pim     {\ensuremath{\pi^-}\xspace}
\def\Km      {\ensuremath{K^-}\xspace}
\def\Dtz     {\ensuremath{D^{*0}_2}\xspace}
\def\Dz      {\ensuremath{D^{*}_0}\xspace}
\def\DoP     {\ensuremath{D_1^\prime}\xspace}
\def\DoPz    {\ensuremath{D_1^{\prime0}}\xspace}
\def\Do      {\ensuremath{D_1}\xspace}
\def\Doz     {\ensuremath{D_1^0}\xspace}
\def\DoPz    {\ensuremath{D_1^{\prime0}}\xspace}
\def\Dt      {\ensuremath{D^{*}_2}\xspace}
\def\Dzz     {\ensuremath{D^{*0}_0}\xspace}
\def\Dv      {\ensuremath{D^{*}_{v}}\xspace}
\def\Bv      {\ensuremath{B^{*}_{v}}\xspace}
\def\eff     {\ensuremath{\epsilon}\xspace}
\def\NLL     {\ensuremath{\mathrm{NLL}}\xspace}
\def\Nsg     {\ensuremath{N_\mathrm{signal}}\xspace}
\def\Nev     {\ensuremath{N_\mathrm{event}}\xspace}
\def\Fbg     {\ensuremath{f_{\mathrm{bg}}}\xspace}
\def\MMDzPma {\ensuremath{m^2(D^+\pi^-_1)}\xspace}
\def\MMDzPmb {\ensuremath{m^2(D^+\pi^-_2)}\xspace}
\def\MMminDP {\ensuremath{m^2_\mathrm{min}(D\pi)}\xspace}
\def\MMmaxDP {\ensuremath{m^2_\mathrm{max}(D\pi)}\xspace}
\def\MMDP    {\ensuremath{m^2(D\pi)}\xspace}
\def\MMPP    {\ensuremath{m^2(\pi\pi)}\xspace}
\def\ct      {\ensuremath{\cos\theta}\xspace}
\def\Rtm     {\ensuremath{{\cal{R}}_{\mathrm{TM}}}\xspace}
\def\Rscf    {\ensuremath{{\cal{R}}_{\mathrm{SCF}}}\xspace}
\def\Fscf    {\ensuremath{f_{\mathrm{SCF}}}\xspace}
\def\Pgen    {\ensuremath{\mathrm{PDF}_{\mathrm{gen}}}\xspace}
\def\Prec    {\ensuremath{\mathrm{PDF}_{\mathrm{acc}}}\xspace}
\def\Ngen    {\ensuremath{N_{\mathrm{gen}}}\xspace}
\def\Nrec    {\ensuremath{N_{\mathrm{acc}}}\xspace}
\def\ea{{\em et al.}}

\def\mev     {\ensuremath{\rm \,Me\kern -0.08em V}\xspace}
\def\mevc    {\ensuremath{{\rm \,Me\kern -0.08em V\!/}c}\xspace}
\def\mevcc   {\ensuremath{{\rm \,Me\kern -0.08em V\!/}c^2}\xspace}
\def\gev     {\ensuremath{\rm \,Ge\kern -0.08em V}\xspace}
\def\gevc    {\ensuremath{{\rm \,Ge\kern -0.08em V\!/}c}\xspace}
\def\gevcc   {\ensuremath{{\rm \,Ge\kern -0.08em V\!/}c^2}\xspace}
\def\gevccs  {\ensuremath{{\rm \,Ge\kern -0.08em V^2\!/}c^4}\xspace}

\newcommand{\dedx}{\ensuremath{\mathrm{d}\hspace{-0.1em}E/\mathrm{d}x}\xspace}
\newcommand{\nimBaseA}       {Nucl.\ Instrum.\ Methods Phys.\ Res., Sect.\ A}
\newcommand{\nima}      [1]  {\nimBaseA~{\bf #1}}

\providecommand{\BDC}{\ensuremath{B^-\to D^{+}\pi^-\pi^-}\xspace}
\providecommand{\DPP}{\ensuremath{D^{+}\pi^-\pi^-}\xspace}
\providecommand{\KPP}{\ensuremath{K^{-}\pi^+\pi^+}\xspace}
\providecommand{\Br}{\ensuremath{\mathcal{B}}\xspace}
\providecommand{\DP}{\ensuremath{D\pi}\xspace}
\providecommand{\DsP}{\ensuremath{D^*\pi}\xspace}
\providecommand{\DzPm}{\ensuremath{D^{+}\pi^-}\xspace}
\providecommand{\Cth}{\ensuremath{\cos\Theta_{th}}\xspace}

\begin{document}

\begin{flushleft}
\babar-PUB-\BABARPubYear/\BABARPubNumber, SLAC-PUB-\SLACPubNumber
\end{flushleft}

\title{\boldmath Dalitz Plot Analysis of \BDC}

%
\author{B.~Aubert}
\author{M.~Bona}
\author{Y.~Karyotakis}
\author{J.~P.~Lees}
\author{V.~Poireau}
\author{E.~Prencipe}
\author{X.~Prudent}
\author{V.~Tisserand}
\affiliation{Laboratoire de Physique des Particules, IN2P3/CNRS et Universit\'e de Savoie, F-74941 Annecy-Le-Vieux, France }
\author{J.~Garra~Tico}
\author{E.~Grauges}
\affiliation{Universitat de Barcelona, Facultat de Fisica, Departament ECM, E-08028 Barcelona, Spain }
% ...
\collaboration{The \babar\ Collaboration}
\noaffiliation

\begin{abstract}
We report on a Dalitz plot analysis of \BDC decays, based on a sample of
about $383 \times 10^6$ $\Y \to \BB$ decays collected with the \babar\ detector
at the \peptwo\ asymmetric-energy \B~Factory at SLAC.
We find the total branching fraction of the three-body decay: 
$\Br(\BDC) = (1.08 \pm 0.03\pm 0.05) \times 10^{-3}$.
We observe the established \Dtz and confirm the existence of \Dzz
in their decays to $D^+\pi^-$,
where the \Dtz and \Dzz are the $2^+$ and $0^+ ~c\bar{u}$ P-wave states, respectively.
We measure the masses and widths of \Dtz and \Dzz to be:
$m_{\Dtz} = (2460.4\pm1.2\pm1.2\pm1.9) \mevcc$,
$\Gamma_{\Dtz} = (41.8\pm2.5\pm2.1\pm2.0) \mev$,
$m_{\Dzz} = (2297\pm8\pm5\pm19) \mevcc$ and $\Gamma_{\Dzz} = (273\pm12\pm17\pm45) \mev$.
The stated errors reflect the statistical and systematic uncertainties,
and the uncertainty related to the assumed composition of signal events
and the theoretical model.
\end{abstract}

\pacs{13.25.Hw, 14.40.Lb, 14.40.Nd}

\maketitle

\section{INTRODUCTION}
\label{sec:Introduction}
                                                                                                                               
Orbitally excited states of the $D$ meson, denoted here as \DJ,
where $J$ is the spin of the meson, provide a unique opportunity to test
the Heavy Quark Effective Theory (HQET)~\cite{hqet1,hqet2}.
The simplest \DJ meson consists of a charm quark and a light anti-quark in
an orbital angular momentum $L=1$ (P-wave) state.
Four such states are expected with spin-parity
$J^P=$ $0^+$ $(j=1/2)$, $1^+$ $(j=1/2)$, $1^+$ $(j=3/2)$ and $2^+$ $(j=3/2)$,
which are labeled here as \Dz, \DoP, \Do and \Dt, respectively,
where $j$ is a quantum number corresponding to the sum of the light quark spin and the orbital
angular momentum $\vec{L}$.

\begin{figure}[tbhp]
\epsfig{width=0.45\textwidth,clip=true,file=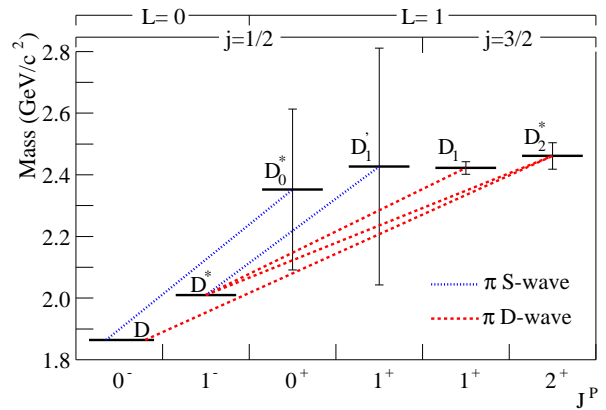}
\caption{Mass spectrum for $c\bar{u}$ states.
The vertical bars show the widths.
Masses and widths are from Ref.~\cite{PDG}.
The dotted and dashed lines between the levels show the dominant pion transitions. 
Although it is not indicated in the figure, 
the two $1^+$ states may be mixtures of $j=1/2$ and $j=3/2$, and
\DoP may decay via a D-wave and \Do may decay via an S-wave.}
\label{fig:lv}
\end{figure}

The conservation of parity and angular momentum in strong interactions imposes constraints
on the strong decays of \DJ states to \DP and \DsP.
The $j=1/2$ states are predicted to decay exclusively through an S-wave:
$\Dz \to \DP$ and $\DoP \to \DsP$.
The $j=3/2$ states are expected to decay through a D-wave:
$\Do \to \DsP$ and $\Dt \to$ \DP and \DsP.
These transitions are summarized in Fig.~\ref{fig:lv}.
Because of the finite $c$-quark mass, the two $J^P=1^+$ states
may be mixtures of the $j=1/2$ and $j=3/2$ states.
Thus the broad \DoP state may decay via a D-wave and
the narrow \Do state may decay via an S-wave.
The $j=1/2$ states with $L=1$, which decay through an
S-wave, are expected to be wide (hundreds of \mevcc),
while the $j=3/2$ states that decay through a D-wave are expected to
be narrow (tens of \mevcc)~\cite{hqet2,falk,falk2}.
Properties of the $L=1$ \DJz mesons~\cite{PDG}
are given in Table~\ref{tab:dstst-properties}.

The narrow \DJ mesons have been previously observed and studied by a number of
experiments~\cite{argus,e691,cleo1,cleo2,e687,cleo3,belle-prd,focus,belle-d1todpp,cdf,belle-d0pp}.
\DJ mesons have also been studied in semileptonic \B decays~\cite{aleph,cleo-semi,D0,delphi,babar-051101,belle-liventsev,babar-0333,babar-0528}. 
Precise knowledge of the properties of the \DJ mesons is important to reduce uncertainties in the measurements
of semileptonic decays, and thus the determination of the Cabibbo-Kobayashi-Maskawa~\cite{CKM} matrix elements $|V_{cb}|$ and $|V_{ub}|$.
The Belle Collaboration has reported the first observation of the
broad \Dzz and \DoPz mesons in \B decay \cite{belle-prd}.
The FOCUS Collaboration has found evidence for broad structures in \DzPm final states~\cite{focus} with
mass and width in agreement with the \Dzz found by Belle Collaboration.
However, the Particle Data Group \cite{PDG} considers that the $J$ and $P$ quantum numbers of
the \Dzz and \DoPz states still need confirmation.

In this analysis, we fully reconstruct the decays \BDC~\cite{conjugate}
and measure their branching fraction.
We also perform an analysis of the Dalitz plot (DP) to measure the exclusive branching fractions of
$\Bm \to \DJz \pim$ and study the properties of the \DJz mesons.
The decay \BDC is expected to be dominated by the
intermediate states $\Dtz\pim$ and $\Dzz\pim$, and has a possible
contribution from \BDC nonresonant (NR) decay.
The \Doz and \DoPz states can not decay strongly
into \DP because of parity and angular momentum conservation.
However, the $D^*(2007)^0$ (labeled as \Dv here) mass is close to the \DP production
threshold and it may contribute as a virtual intermediate state.
The {\ensuremath{B^{*}}} (labeled as \Bv here) produced in a virtual process $B^-\to \Bv \pi^-$ may also contribute via
the decay ~$\Bv \to \DzPm$.
Possible contributions from these virtual states are also studied in this analysis.

\begin{table}[tbhp]
  \begin{center}
  \caption{Properties of $L=1$ \DJz mesons~\cite{PDG}.}
  \begin{tabular}{|llcccc|}
    \hline
    ~& $J^P$ & Mass & Width & Decays   & Partial \\
    ~&       & (\mevcc)     & (\mev) &  seen~\cite{PDG} & waves \\
    \hline
     \Dzz      & $0^+$ & $2352\pm 50$ & $261\pm50$ & \DP & S \\ 
     \DoPz     & $1^+$ & $2427\pm36 $ & $384^{+130}_{-105}$ & \DsP & S, D \\ 
     \Doz      & $1^+$ & $2422.3\pm 1.3$& $20.4\pm1.7$ & \DsP, $D^0\pi^+\pi^-$ & S, D \\ 
     \Dtz      & $2^+$ & $2461.1\pm 1.6$& $43\pm 4$ & \DsP, \DP & D \\ 
    \hline
  \end{tabular}
  \label{tab:dstst-properties}
  \end{center}
\end{table}

\section{THE \babar\ DETECTOR AND DATASET}
\label{sec:DetectorAndData}

The data used in this analysis were collected with the \babar\ detector
at the \peptwo\ asymmetric-energy \epem\ storage rings at SLAC between 1999 and
2006. The sample consists of 347.2 \invfb\ corresponding
to $(382.9\pm4.2) \times 10^6$ $\BB$ pairs ($N_{\BB}$) taken on the peak of the \Y resonance.
Monte Carlo (MC) simulation is used to study the detector response, its acceptance, background,
and to validate the analysis.
We use GEANT4~\cite{geant4} to simulate resonant $\epem \to \Y \to \BB$ events
(generated by EvtGen~\cite{evtgen}) and $\epem \to \qqbar$ (where $q=u,d,s$ or $c$) continuum events
(generated by JETSET~\cite{jetset}).

A detailed description of the \babar\ detector is given in Ref.~\cite{babarNim}. 
Charged particle trajectories are measured by a five-layer, double-sided
silicon vertex tracker (SVT) and a 40-layer drift chamber (DCH) immersed
in a 1.5~T magnetic field. Charged particle identification (PID) is achieved
by combining information from a ring-imaging Cherenkov device with
ionization energy loss (\dedx) measurements in the DCH and SVT.

\section{EVENT SELECTION}
\label{sec:EventSelection}

\begin{figure}[tb]
\epsfig{width=0.45\textwidth,clip=true,file=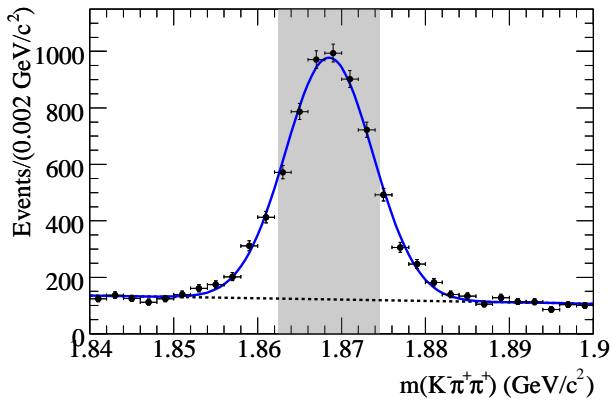}
\caption{$K^+\pi^-\pi^-$ invariant mass distribution for \Dp candidates for the selected \BDC decays
without the cut on the mass of \Dp.
Data (points with statistical errors) are compared to the results of the fit (solid curve),
with the background distribution marked as a dashed line.
The shaded area marks the \Dp signal region.}
\label{fig:MassD}
\end{figure}

Five charged particles are selected to reconstruct decays of \BDC with $D^+\to K^-\pi^+\pi^+$.
The charged particle candidates are required to have 
transverse momenta above $100 \mevc$ and at least twelve hits in the DCH.
A \Km candidate must be identified as a kaon using a likelihood-based particle identification
algorithm (with an average efficiency of $\sim$85\% and an average misidentification probability of $\sim$3\%).
Any combination of $K^-\pi^+\pi^+$ candidates with a common vertex and an invariant
mass between 1.8625 and 1.8745 \gevcc is accepted as a \Dp candidate.
We fit the invariant mass distribution of the $K^-\pi^+\pi^+$ candidates
with a function that includes a Gaussian component for the
signal and a linear term for the background.
The signal parameters (mean and width of Gaussian) and slope of the background function are free parameters of the fit.
The data and the result of the fit are shown in Fig.~\ref{fig:MassD}.
The invariant mass resolution for this \Dp decay is about 5.2 \mevcc.
The \Bm candidates are reconstructed by combining a \Dp candidate
and two charged tracks. The trajectories of the three daughters of
the \Bm meson candidate are constrained to originate from a common decay vertex.
The \Dp and \Bm vertex fits are required to have converged.

At the \Y resonance, \B mesons can be characterized by two nearly independent kinematic variables,
the beam-energy substituted mass \mes and the energy difference \de:
\begin{eqnarray}
\label{eq:mesde}
\mes&=&\sqrt{(s/2+\vec{p}_0 \cdot \vec{p}_B)^2/E_0^2-p_B^2}, \\
\Delta E&=&E_B^*-\sqrt{s}/2,
\end{eqnarray}
where $E$ and $p$ are energy and momentum, the
subscripts 0 and $B$ refer to the \epem-beam system and the \B candidate, respectively;
$s$ is the square of the center-of-mass energy and the asterisk 
labels the center-of-mass frame. 
For \BDC signal events,
the \mes distribution is well described by a Gaussian resolution function
with a width of 2.6 \mevcc centered at the \Bm meson mass,
while the \de distribution can be represented by a sum of two Gaussian functions with a
common mean near zero and different widths with a combined RMS of 20 MeV.

Continuum events are the dominant background.
Suppression of background from continuum events is provided by two topological requirements.
In particular, we employ restrictions on the magnitude of the cosine of the thrust angle, \Cth, defined as the angle between
the thrust axis of the selected \B candidate and the thrust axis of the remaining tracks and neutral clusters in the event.
The distribution of $|\Cth|$ is strongly peaked towards unity for continuum background
but is uniform for signal events.
We also select on the ratio of the second to the zeroth Fox-Wolfram moment~\cite{r2}, $R_2$, to
further reduce the continuum background.
The value of $R_2$ ranges from 0 to 1.
Small values of $R_2$ indicate a more spherical event shape (typical for a \BB event) 
while values close to 1 indicate a 2-jet event topology (typical for a \qqbar event).
We accept the events with $|\Cth| < 0.85$ and $R_2 < 0.30$.
The $|\Cth|$ ($R_2$) cut eliminates about 68\% (71\%) of the continuum background while retaining about
90\% (83\%) of signal events.

\begin{figure}[tb]
\epsfig{width=0.45\textwidth,clip=true,file=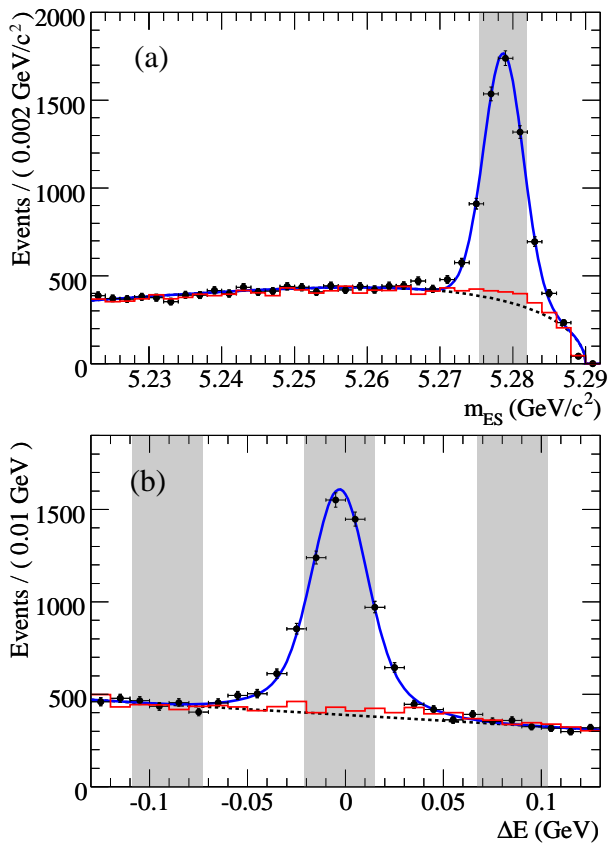}
\caption{(a) \mes and (b) \de distributions for $D^+\pi^-\pi^-$ candidates.
Data (points with statistical errors) are compared to the results of the fits (solid curves),
with the background contributions marked as dashed lines.
The histograms are the corresponding distributions of the background MC sample as described in the text.
The shaded area in (a) shows the signal region,
while the three shaded areas in (b) mark the signal region in the center and the two sidebands.}
\label{fig:dataDeltaE}
\end{figure}

To suppress backgrounds, restrictions are placed on \mes: $5.2754<\mes<5.2820$ \gevcc,
and \de: $-130<\de<130$ \mev.
The selected samples of \B candidates are used as input to an unbinned
extended maximum likelihood fit to the \de distribution.
The result of the fit is used to determine the fractions of signal and background events in the selected data sample.
For events with multiple candidates ($\sim3.5$\% of the selected events) satisfying the selection criteria,
we choose the one with best $\chi^2$ from the \B\-vertex fit.
Based on MC simulation, we determine that the correct candidate is selected at least 65\% of the time.
We fit the \mes distribution of the selected \BDC candidates with a sum of a Gaussian function for the signal
and a background function for the background having the probability density,
$P(x) \propto x\sqrt{1-x^2}\exp(-\xi(1-x^2))$, where $x=\mes/m_0$ with $m_0$ fixed at 5.29 \gevcc and
$\xi$ is a shape parameter~\cite{argus2}.
The signal parameters (mean, width of Gaussian) and the shape parameter of the background function are free parameters of the fit.
The data and the result of the fit are shown in Fig.~\ref{fig:dataDeltaE}a.
We fit the \de distribution of the selected \BDC candidates with a sum of two Gaussian functions with a common mean for the
signal and a linear function for the background.
The signal parameters (mean, width of wide Gaussian, width and fraction of narrow Gaussian)
and the slope of the background function are free parameters of the fit.
The data and the result of the fit are shown in Fig.~\ref{fig:dataDeltaE}b. 
The resulting signal yield is $3496\pm 74$ events, where the error is statistical only.
A clear signal is evident in both \mes and \de distributions.

To distinguish signal and background in the Dalitz plot studies,
we divide the candidates into three subsamples:
the signal region, $-21<\de<15\mev$, the left sideband, $-109<\de<-73\mev$,
and the right sideband, $67<\de<103\mev$.
The events in the signal region are used in the Dalitz plot analysis, while
the events in the sideband regions are used to study the background.

In order to check the shape of the background \de distribution, we have generated a background MC sample
of resonant and continuum events with \BDC signal events removed.
The background MC sample has been scaled to the same luminosity as the data.
The \de distribution of the selected events from the background MC sample is
shown as the histogram in Fig.~\ref{fig:dataDeltaE}b.
A small amount of peaking background is found from misreconstructed decays of
$\Bzb \to D^+\rho^-$ with $\rho^-\to\pi^-\pi^0$,
where a $\pi^0$ is missed and a random track in the event is misidentified as a signal $\pi^-$.
The background histogram in Fig.~\ref{fig:dataDeltaE}b is fitted with a sum of two Gaussian functions with a common mean for the
peaking background, with parameters fixed to those obtained from the fit to data, and a linear function
to describe the combinatorial background. The amount of peaking background is estimated at $82\pm 41$ events.
After peaking background subtraction, the number of signal events above background is $\Nsg=3414\pm85$.
The background fraction in the signal region is $(30.4\pm1.1) \%$.

\section{Dalitz Plot Analysis}
\label{sec:dalitzPlotAnalysis}

We refit the \Dp and \Bm candidate momenta by
constraining the trajectories of the three daughters of the \Bm meson candidate to originate from a common decay vertex
while constraining the invariant masses of \KPP and \DPP to the \Dp and \Bm masses~\cite{PDG}, respectively.
The mass-constraints ensure that all events fall within the Dalitz plot boundary.

In the decay of a \Bm into
a final state composed of three pseudo-scalar particles (\DPP), two
degrees of freedom are required to describe the decay kinematics.
In this analysis we choose the two \DP invariant mass-squared combinations $x=\MMDzPma$
and $y=\MMDzPmb$ as the independent variables, where the two like-sign pions $\pi^-_1$ and $\pi^-_2$ are randomly assigned to $x$ and $y$.
This has no effect on our analysis since the likelihood
function (described below) is explicitly symmetrized with respect
to interchange of the two identical particles.

The differential decay rate is generally given in terms of the Lorentz-invariant matrix element $\ensuremath{\mathcal{M}}$ by
\begin{eqnarray}
\label{dalitz}
\frac{d^2\Gamma}{dx dy}=\frac{| \ensuremath{\mathcal{M}} |^2}{256\pi^3m_B^3},
\end{eqnarray}
where $m_B$ is the \B meson mass.
The Dalitz plot gives a graphical representation of the variation of the square
of the matrix element, $|\ensuremath{\mathcal{M}}|^2$, over the kinematically accessible phase space ($x$,$y$) of the
process. Non-uniformity in the Dalitz plot can indicate presence of intermediate
resonances, and their masses and spin quantum numbers can be determined.

\subsection{Probability Density Function}

We describe the distribution of candidate events in the Dalitz
plot in terms of a probability density function (PDF).
The PDF is the sum of signal and background components and has the form:
\begin{eqnarray}
\mathrm{PDF}(x,y) &=& \Fbg \frac{B(x,y)}{\int_{\mbox{DP}} B(x,y) dxdy} \nonumber \\
&+& (1-\Fbg) \frac{ \left [ S(x,y) \otimes {\cal{R}}\right ] \eff (x,y)}{ \int_{\mbox{DP}} \left [ S (x,y) \otimes {\cal{R}}\right ] \eff (x,y) dxdy}, \nonumber \\
\label{pdfall}
\end{eqnarray}
where the integral is performed over the whole Dalitz plot,
the $S(x,y) \otimes {\cal{R}}$ is the signal term convolved with the signal resolution function,
$B(x,y)$ is the background term,
\Fbg is the fraction of background events, and
\eff is the reconstruction efficiency.

An unbinned maximum likelihood fit to the Dalitz plot is performed in order to maximize the value of
\begin{eqnarray}
{\cal L}= \prod_{i=1}^{\Nev} \mathrm{PDF}(x_i,y_i)
\label{eq:likelihood}
\end{eqnarray}
with respect to the parameters used to describe $S$,
where $x_i$ and $y_i$ are the values of $x$ and $y$ for event $i$ respectively,
and \Nev is the number of events in the Dalitz plot.
In practice, the negative-log-likelihood (\NLL) value
\begin{eqnarray}
\mathrm{NLL}=-\ln{\cal L}
\end{eqnarray}
is minimized in the fit.

\subsection{Goodness-of-fit}
It is difficult to find a proper binning at the kinematic
boundaries in the $x$-$y$ plane of the Dalitz plot. For this reason, we choose to
estimate the goodness-of-fit $\chi^2$ in the \ct (range from -1 to 1)
and \MMminDP (range from 4.04 to 15.23 \gevccs) plane,
which is a rectangular representation of the Dalitz plot.
The parameter $\theta$ is the helicity angle of the $D\pi$ system
and \MMminDP is the lesser of $x$ and $y$.
The helicity angle $\theta$ is defined as the angle between the momentum vector of the pion
from the $B$ decay (bachelor pion)
and that of the pion of the $D\pi$ system in the $D\pi$ rest-frame.

The $\chi^2$ value is calculated using the formula
\begin{eqnarray}
\chi^2 = \sum_{i}\chi_i^2 = \sum_{i=1}^{n_\mathrm{total}} \frac{ ({N_\mathrm{cell}}_i - {N_\mathrm{fit}}_i)^2}{{N_\mathrm{fit}}_i},
\label{chi2}
\end{eqnarray}
for cells in a $18\times18$ grid of the two-dimensional histogram.
In Eq.~(\ref{chi2}), $n_\mathrm{total}$
is the total number of cells used, ${N_\mathrm{cell}}_i$ is the number of events in each cell,
${N_\mathrm{fit}}_i$ is the expected number of events in that cell as
predicted by the fit results.
The number of degrees of freedom (NDF) is calculated as $n_\mathrm{total}-k-1$, where
$k$ is the number of free parameters in the fit.
We require ${N_\mathrm{fit}}\ge10$;
if this requirement is not
met then neighboring cells are combined until ten events are accumulated.

\subsection{Matrix element $\ensuremath{\mathcal{M}}$ and Fit Parameters}
This analysis uses an isobar model formulation in which the signal decays
are described by a coherent sum of
a number of two-body ($D\pi$ system + bachelor pion) amplitudes.
The orbital angular momentum between the $D\pi$ system and the bachelor
pion is denoted here as $L$.
The total decay matrix element $\ensuremath{\mathcal{M}}$ for $\BDC$ is given by:
\begin{eqnarray}
\ensuremath{\mathcal{M}} &=& \sum_{L=(0,1,2)} \rho_{L}e^{i\phi_{L}}\left [ N_L(x,y) + N_L(y,x) \right ] \nonumber \\
                         &+& \sum_k                  \rho_{k}e^{i\phi_{k}}\left [ A_k(x,y) + A_k(y,x) \right ], 
\end{eqnarray}
where the first term represents the S-wave ($L=0$), P-wave ($L=1$) and D-wave ($L=2$) nonresonant contributions,
the second term stands for the resonant contributions,
the parameters $\rho_{k}$ and $\phi_{k}$ are the magnitudes and phases of $k^{\rm th}$ resonance,
while $\rho_{L}$ and $\phi_{L}$ correspond to the magnitudes and phases of the nonresonant contributions with angular momentum $L$.
The functions $N_L(x,y)$ and $A_k(x,y)$ are the amplitudes for nonresonant and resonant terms, respectively.

The resonant amplitudes $A_k(x,y)$ are expressed as:
\begin{eqnarray}
A_k(x,y) = R_k(m) F_L(p^\prime r^\prime) F_L(qr) T_L(p,q,\ct),
\label{ampres}
\end{eqnarray}
where $R_k(m)$ is the $k^{\rm th}$ resonance lineshape,
$F_L(p^\prime r^\prime)$ and $F_L(qr)$ are the Blatt-Weisskopf barrier factors~\cite{formf},
and $T_L(p,q,\ct)$ gives the angular distribution.  The parameter $m$ ($=\sqrt{x}$) is the invariant mass of the $D\pi$ system.
The parameter $p^\prime$ is the magnitude of the three momentum of the bachelor pion evaluated in the $B$-meson rest frame.
The parameters $p$ and $q$ are the magnitudes of the three momenta of the bachelor pion and the pion of the
$D\pi$ system, both in the $D\pi$ rest frame.
The parameters $p^\prime$, $p$, $q$ and $\theta$ are functions of $x$ and $y$.

The nonresonant amplitudes $N_L(x,y)$ with $L=0,1,2$ are similar to $A_k(x,y)$ but do not contain resonant mass terms:
\begin{eqnarray}
N_0(x,y) &=& 1, \\
N_1(x,y) &=& F_1(p^\prime r^\prime) F_1(qr) T_1(p,q,\ct),\\
N_2(x,y) &=& F_2(p^\prime r^\prime) F_2(qr) T_2(p,q,\ct).
\label{ampnr}
\end{eqnarray}

The Blatt-Weisskopf barrier factors $ F_L(p^\prime r^\prime)$ and $F_L(qr)$
depend on a single parameter, $r^\prime$ or $r$, the radius of the barrier, which we take to be $1.6~\mathrm{(GeV/c)}^{-1}$,
similarly to Ref.~\cite{belle-prd}.
A discussion of the systematic uncertainty associated with the choice of the values of $r$ and $r^\prime$ follows below.
The forms of $F_L(z)$, where $z = p^\prime r^\prime$ or $qr$, for $L=0, 1, 2$ are:
\begin{eqnarray}
F_0(z) &=& 1,\\
F_1(z) &=& \sqrt{\frac{1+z^2_0}{1+z^2}},\label{barrier1}\\
F_2(z) &=& \sqrt{\frac{9+3z^2_0+z^4_0}{9+3z^2+z^4}},
\label{barrier2}
\end{eqnarray}
where $z_0 = p^\prime_0r^\prime$ or $q_0r$.
Here $p^\prime_0$ and $q_0$ represent the values of $p^\prime$ and $q$, respectively,
when the invariant mass is equal to the pole mass of the resonance.
For nonresonant terms,
the fit results are not affected by the choice of invariant mass (we use the sum of $m_D$ and $m_\pi$) used for the
calculations of $p^\prime_0$ and $q_0$.
For virtual \Dv decay, $\Dv\to D^+\pi^-$, and virtual \Bv production in $\B^- \to \Bv \pi^-$,
we use an exponential form factor in place of the Blatt-Weisskopf barrier factor, as discussed in Ref.~\cite{belle-prd}:
\begin{eqnarray}
F(z) = \exp{(-(z-z^\prime))},
\label{barrier3}
\end{eqnarray}
where $z^\prime=r p_v$ for $\Dv\to D^+\pi^-$ and $z^\prime=r^\prime p_v$ for $\B^- \to \Bv \pi^-$.
Here, we set $p_v=0.038~\mathrm{GeV/c}$, which gives the best fit,
although any value of $p_v$ between 0.015 and $1.5~\mathrm{GeV/c}$ gives negligible effect
on the fitted parameters compared to their statistical errors.

The resonance mass term $R_k(m)$ describes the intermediate resonance.
All resonances in this analysis are parametrized with relativistic Breit-Wigner functions:
\begin{eqnarray}
      R_k(m) = \frac{1}{(m_0^2-m^2)-im_0\Gamma(m)},
\end{eqnarray}
where the decay width of the resonance depends on $m$:
\begin{eqnarray}
\Gamma(m) = \Gamma_0 \left ( \frac{q}{q_0} \right ) ^{2L+1} \left ( \frac{m_0}{m} \right ) F^2_L(qr),
\end{eqnarray}
where $m_0$ and $\Gamma_0$ are the values of the resonance pole mass and decay width, respectively.

The terms $T_L(p,q,\ct)$ describe the angular distribution of final state particles and
are based on the Zemach tensor formalism~\cite{Zemach}.
The definitions of $T_L(p,q,\ct)$ for $L=0,1,2$ are:
\begin{eqnarray}
T_0(p,q,\ct) &=& 1,\\
T_1(p,q,\ct) &=& -2pq\ct,\\
T_2(p,q,\ct) &=& 4p^2q^2(\cos^2 \theta-1/3).
\end{eqnarray}

The signal function is then given by:
\begin{eqnarray}
 S(x,y) = | \ensuremath{\mathcal{M}}|^2.
\label{signalfunction}
\end{eqnarray}

In this analysis, 
the masses of \Dv and \Bv are taken from the world averages~\cite{PDG} while their widths are fixed at 0.1 \mev;
the magnitude $\rho_{k}$ and phase $\phi_{k}$ of the \Dtz amplitude are fixed to 1 and 0, respectively,
while the masses and widths of \DJz resonances and other magnitudes and phases are free parameters to be determined in the fit.
The effect of varying the masses of \Dv and \Bv within 
their errors~\cite{PDG} and widths of \Dv and \Bv between 0.001 and 0.3 \mev
is negligible compared to the other model-dependent systematic uncertainties given below.

Since the choice of normalization, phase convention and amplitude
formalism may not always be the same for different experiments, we use fit fractions and relative phases
instead of amplitudes to
allow for a more meaningful comparison of results. The fit fraction for
the $k^{\rm th}$ decay mode is defined as the integral of the resonance decay
amplitudes divided by the coherent matrix element squared for the
complete Dalitz plot:
\begin{eqnarray}
f_k = \frac{\int_{\mbox{DP}} | \rho_{k}(A_k(x,y)+A_k(y,x)) |^2 dxdy} {\int_{\mbox{DP}} |\mathcal{M} |^2 dxdy}.
\label{equ:fitFraction}
\end{eqnarray}
The fit fraction for nonresonant term with angular momentum $L$ has a similar form:
\begin{eqnarray}
f_L = \frac{\int_{\mbox{DP}} | \rho_{L}(N_L(x,y)+N_L(y,x)) |^2 dxdy} {\int_{\mbox{DP}} |\mathcal{M} |^2 dxdy}.
\label{equ:fitFraction1}
\end{eqnarray}
The fit fractions do not necessarily add up to unity because of interference among the amplitudes.

To estimate the statistical uncertainties on the fit fractions,
the fit results are randomly modified according to the covariance matrix of the fit 
and the new fractions are computed using Eq.~(\ref{equ:fitFraction}) or (\ref{equ:fitFraction1}).
The resulting fit fraction distribution is fitted with a Gaussian whose width gives
the error on the given fraction.

\subsection{Signal Resolution Function}
The detector has finite resolution, thus measured quantities differ from their true values.
For the narrow resonance $D^*_2$ with the expected width of about 40 MeV,
the signal resolution needs to be taken into account.
In order to obtain the signal resolution on $m^2(D\pi)$ around the $D^*_2$ mass region, we study a sample of MC generated
$B^-\to X \pi^- \to D^+\pi^-\pi^-$ decays, with the mass and width of $X$ set to 2.460 \gevcc ($D^*_2$ mass region)
and 0 MeV, respectively,
and subject these events to the same analysis reconstruction chain.
The reconstructed events are then classified into two categories:
truth-matched (TM) events, where the $B$ and the daughters are correctly reconstructed,
and self-crossfeed (SCF) events, where one or more of the daughters is not correctly associated with the generated particle.

\begin{figure}[tbp]
\epsfig{width=0.45\textwidth,clip=true,file=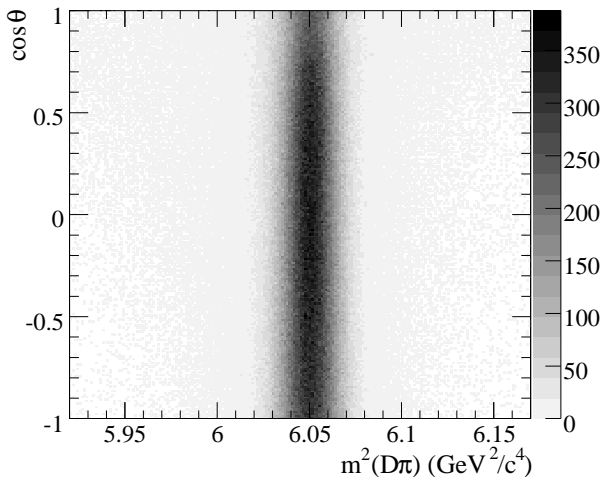}
\caption{Two-dimensional histogram \ct versus \MMDP of the
truth-matched events as defined in the text.}
\label{fig:TMresolution}
\end{figure}

The two-dimensional distribution of \ct versus \MMDP for truth-matched
events is shown in Fig.~\ref{fig:TMresolution}. 
Since the resolution is independent of $\cos\theta$,
we fit the distribution of the quantity  $q^\prime =\MMDP - m^2_\mathrm{true}$
using a sum of two Gaussian functions with a common mean
to obtain the resolution function for truth-matched events (\Rtm).
The signal resolution for
an invariant mass of the \DP combination around the \Dtz region is about 3 \mevcc.

\begin{figure}[tbp]
\epsfig{width=0.45\textwidth,clip=true,file=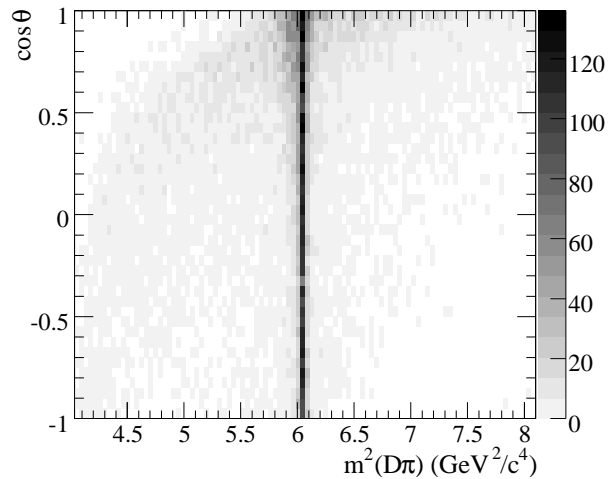}
\caption{Two-dimensional histogram \ct versus \MMDP of the
self-crossfeed events as defined in the text.}
\label{fig:SCFresolution}
\end{figure}

\begin{figure}[tbp]
\centering \epsfig{width=0.45\textwidth,file=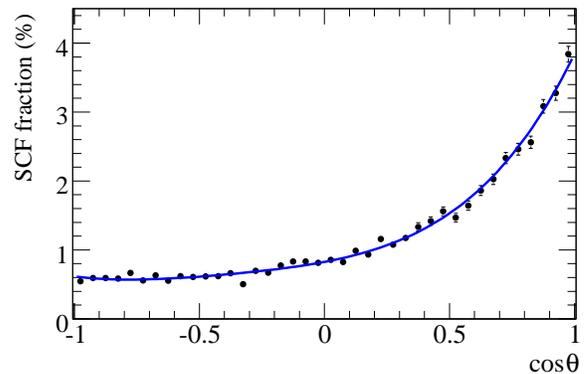}
\caption{$\Fscf(\ct)$ distribution. 
The observed self-crossfeed fractions (points with statistical errors) are compared to the results of the fit (solid curve).}
\label{fig:Fscf}
\end{figure}

The two-dimensional distribution of \ct versus \MMDP for
self-crossfeed events is shown in Fig.~\ref{fig:SCFresolution}.
The SCF fraction, $\Fscf$, varies from 0.5\% to 4.0\% with \ct.
We fit the $\Fscf$ distribution with a 4th-order polynomial function.
The $\Fscf$ distribution and the result of the fit are shown in Fig.~\ref{fig:Fscf}.
The resolution for self-crossfeed events varies between 5 \mevcc and 100 \mevcc with \ct.
We divide the \ct interval into 40 bins of equal width and use these
bins to describe the resolution function (\Rscf) in terms of a sum of 
two bifurcated Gaussian (BGaussian) functions with different means.
The BGaussian is a Gaussian as a function
of $q^\prime$ with three parameters, $q^\prime_0$ the mean, and the two
widths, $\sigma_1$ on the left and $\sigma_2$ on the right side of the mean.
The form of BGaussian is:
\begin{eqnarray}
{\mbox{BGaussian}}(q^\prime-q^\prime_0, \sigma_1, \sigma_2) = ~~~~~~~~~~~~~~~~~~~~~~~~~~~~~~\nonumber \\
              \left\{
              \begin{array}{ll}
                   \frac{2}{\sqrt{2\pi}(\sigma_1+\sigma_2)}
                   \exp(-\frac{(q^\prime-q^\prime_0)^2}{2\sigma_1^2}) &  \text{if $q^\prime<q^\prime_0$;}\\
                   \frac{2}{\sqrt{2\pi}(\sigma_1+\sigma_2)}
                   \exp(-\frac{(q^\prime-q^\prime_0)^2}{2\sigma_2^2}) & \text{if $q^\prime\geq q^\prime_0$,}
              \end{array}
       \right.
\label{bgau}
\end{eqnarray}
where $q^\prime_0$, $\sigma_1$ and $\sigma_2$ are free parameters.

The signal resolution function is then given by:
\begin{eqnarray}
{\cal{R}}(q^\prime,\ct) &=& (1-\Fscf (\ct)) \times \Rtm (q^\prime) \nonumber \\
                      &+& \Fscf (\ct) \times \Rscf (q^\prime,\ct).
\label{resolution}
\end{eqnarray}
The function ${\cal{R}}(q^\prime,\ct)$ represents the probability density
for an event having the true mass squared $m^2_\mathrm{true}$ to be
reconstructed at \MMDP for different $\cos\theta$ regions.

The signal term $S$ in Eq.~(\ref{pdfall}) is convoluted with the above resolution function.
For each event, the convolution is performed using numerical integration:
\begin{eqnarray}
S(x,y) \otimes {\cal{R}} = \int S(q_\mathrm{min}+q^\prime, q_\mathrm{max}^{\prime})\times
{\cal{R}}(q^\prime,\ct)  dq^\prime,~~
\label{integration}
\end{eqnarray}
where $S$ is the signal function in Eq.~(\ref{signalfunction}), and $q_\mathrm{min}$ ($q_\mathrm{max}$)
is the lesser (greater) of $x$ and $y$.
The quantity $\ct$ is determined from $q_\mathrm{min}$ and $q_\mathrm{max}$ and is assumed to be
constant during convolution.
The resolution in $\ct$ has a negligible effect on the fitted parameters. 
The quantity $q_\mathrm{max}^{\prime}$ is computed using the kinematics
of three-body decay with $q_\mathrm{min}$, $q^\prime$ and $\ct$.

The resolution function and the integration method in Eq.~(\ref{integration}) have been
fully tested using 262 MC samples with full event reconstruction given below.
We have compared $D$ invariant mass resolutions for $D^0\to K^-\pi^+$, $K^-\pi^+\pi^-\pi^+$
and $D^+\to K^-\pi^+\pi^+$ between data and MC-simulated events 
and find that they agree within their statistical uncertainties.
Estimated biases in the fitted parameters due to 
uncertainties in the signal resolution function are small and have 
been included into the systematic errors.

\subsection{Efficiency}
The signal term $S$ defined above is modified in order to take into account experimental particle 
detection and event reconstruction efficiency.
Since different regions of the Dalitz plot correspond to different event topologies, the efficiency is not
expected to be uniform over the Dalitz plot.
The term $\epsilon(x,y)$ in Eq.~(\ref{pdfall}) is the overall efficiency for truth-matched and self-crossfeed signal events,
hence the efficiency for truth-matched signal events is
\begin{eqnarray}
\epsilon_{\mathrm{TM}}(x,y) = \epsilon(x,y)
(1-f_{\mathrm{SCF}}(\ct)).
\end{eqnarray}

In order to determine the efficiency across the Dalitz plot,
a sample of simulated \BDC events in the Dalitz plot is generated.
Some events are generated with one or more additional final-state photons
to account for radiative corrections~\cite{radiative}.
As a result, the generated Dalitz plot is slightly distorted from the uniform distribution.
The number of generated events is
$N_\mathrm{gen}=$1252k. Each event is subjected to
the standard reconstruction and selection, described in Section III.
In addition, we require that the candidate decay is truth matched.
After correcting for data/MC efficiency differences in particle identification,
which are momentum dependent and thus vary over the Dalitz plot,
the total number of accepted events is
$N_\mathrm{acc} = 121,390$. We employ an unbinned likelihood method
to fit the Dalitz plot distributions for generated and accepted event samples.
The PDF for generated events (\Pgen) is a fourth-order two-dimensional polynomial
while the PDF for accepted events (\Prec) is a seventh-order
two-dimensional polynomial. The efficiency function is then given by:
\begin{eqnarray}
\epsilon_\mathrm{TM}(x,y) = \frac{\Prec(x,y) \times \Nrec }{ \Pgen(x,y) \times \Ngen}.
\label{equ:effpdf}
\end{eqnarray}

\begin{figure}[bt]
\centering \epsfig{width=0.45\textwidth,file=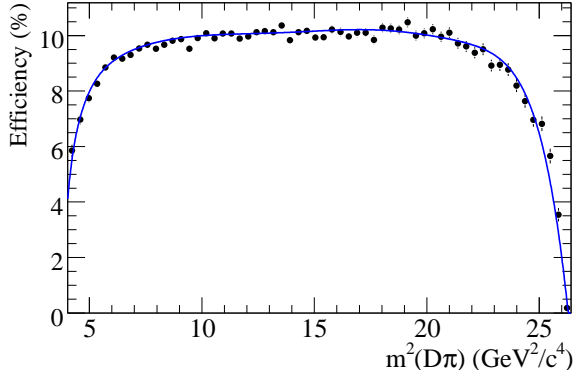}
\caption{The efficiency for signal decays as a function of \MMDP, as determined by MC   
        simulation (points with statistical errors) and the results of the fit to the
        accepted and generated distributions (solid curve).}
\label{fig:effprojection}
\end{figure}

Fig.~\ref{fig:effprojection} shows the efficiency as a function of \MMDP
and the fit result for MC-simulated events.

\subsection{Background}
\label{sec:bgpdf}

\begin{figure}[tb]
\epsfig{width=0.45\textwidth,clip=true,file=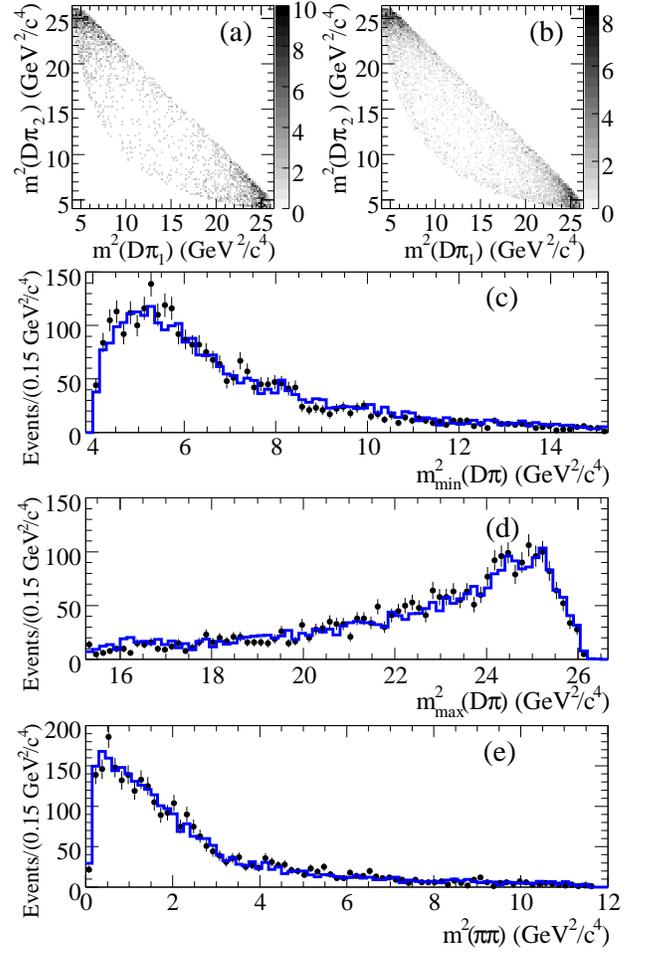}
\caption{Comparison of events in the \de sideband:
Dalitz plot for (a) data and (b) MC-simulated events,
and projections on (c) \MMminDP, (d) \MMmaxDP and
(e) ${m}^2(\pi\pi)$ with data (points with statistical errors) and MC predictions (histograms).}
\label{fig:background}
\end{figure}

\begin{figure}[tb]
\epsfig{width=0.45\textwidth,clip=true,file=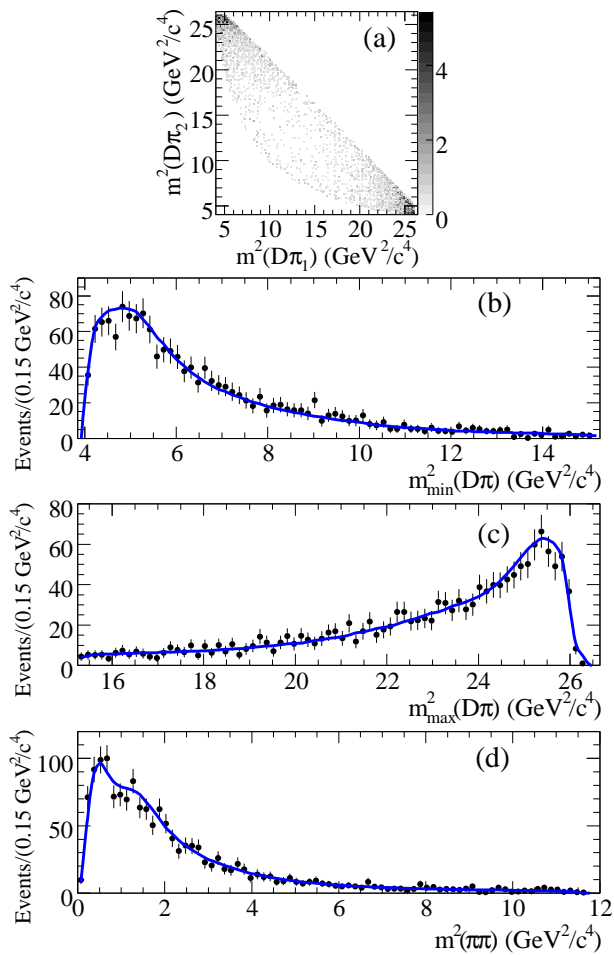}
\caption{Fit to background events in the \de signal region of the MC sample:
(a) Dalitz plot and projections on (b) \MMminDP, (c) \MMmaxDP and
(d) ${m}^2(\pi\pi)$ with MC predictions (points with statistical errors) and
the fits (solid curves).}
\label{fig:fitbackground}
\end{figure}

The background distribution is modeled using
MC background events, selected with the same criteria
applied to the data and requiring the $B$ candidate to fall into the signal
\de region defined in Section III.
Events in the data \de
sidebands could also be used to model the background, however in MC
studies we find differences between the Dalitz plot distributions of the
background in the signal and sideband regions.  Since we find the Dalitz
plot distributions of sideband events in data and in MC simulation to be
consistent within their statistics, we are confident that the MC simulation can accurately represent the
background distribution in the signal region.
Fig.~\ref{fig:background}a, \ref{fig:background}b and Fig.~\ref{fig:fitbackground}a show the
Dalitz plot distributions of sideband events in data, sideband events in the MC sample
and background events in the \de signal region of the MC sample, respectively.
Fig.~\ref{fig:background}c, \ref{fig:background}d and \ref{fig:background}e show the comparisons
of \de sideband events between data and MC simulation in
$\MMminDP$, $\MMmaxDP$ and ${m}^2(\pi\pi)$ projections, respectively.
Here $\MMminDP$ ($\MMmaxDP$) is the lesser (greater) of $x$ and $y$.

The parameterization used to describe the background is:
\begin{eqnarray}
B(x,y)  &=& c_0 (q_\mathrm{min}-q_1)^{c_1} \times  \nonumber\\ 
        &~& \exp{(c_2(q_\mathrm{min}-q_1)+c_3(q_\mathrm{min}-q_1)^2)} \nonumber\\
        &+& c_4 (q_2-q_\mathrm{max})^{c_5} \times \nonumber\\ 
        &~& \exp{(c_6(q_2-q_\mathrm{max})+c_7(q_2-q_\mathrm{max})^2)} \nonumber\\
        &+& c_8 (z-z_1))^{c_9} \exp{(c_{10}(z-z_1)+c_{11}(z-z_1)^2)}\nonumber\\
        &+& c_{15} \mathrm{BGaussian}(q_\mathrm{max}-c_{12},c_{13},c_{14})\nonumber\\
        &+& c_{19} \mathrm{BGaussian}(z-c_{16}, c_{17},c_{18}),
\label{equ:bgfit2}
\end{eqnarray}
where the coefficients $c_0$ to $c_{19}$ are free parameters to be
determined from the fit, $q_1 = (m_D+m_\pi)^2 = 4.04$ $\mbox{GeV}^2/c^4$ and
$q_2 = (m_B-m_\pi)^2=26.41$ $\mbox{GeV}^2/c^4$ are the lower and upper limits of the
Dalitz plot, respectively, $z_1 = (2m_\pi)^2 = 0.077$ $\mbox{GeV}^2/c^4$ is the lower
limit of $m^2(\pi\pi)$, $q_\mathrm{min}$ is the lesser of $x$ and
$y$, $q_\mathrm{max}$ is the greater of $x$ and $y$, $z$ is the
invariant $m^2(\pi\pi)$, and $\mathrm{BGaussian}$ is given in Eq.~(\ref{bgau}).

The projections on $\MMminDP$, $\MMmaxDP$ and ${m}^2(\pi\pi)$ and the 
result of the fit for the background events in the signal region of the MC sample
are shown in Fig.~\ref{fig:fitbackground}b, \ref{fig:fitbackground}c
and \ref{fig:fitbackground}d.
The $\chi^2/\mbox{NDF}$ for the fit is $72/64$.

\section{Results}
\label{sec:Results}

\subsection{Branching Fraction $\Br(\Bm \to D^{+}\pim\pim)$}

The total \BDC branching fraction is calculated using the relation:
\begin{eqnarray}
\Br = \frac{\Nsg}{(\bar{\epsilon} \cdot \Br(D^+)) \cdot 2N(\BpBm)},
\end{eqnarray}
where $\Nsg=3414\pm85$ is the fitted signal yield given in Section~\ref{sec:EventSelection},
$\bar{\epsilon}$ is the average efficiency,
$\Br(D^+) = (9.22\pm0.21)$\% is the branching fraction for $\Dp \to \KPP$~\cite{PDG,cleodpp},
and the total number of \BpBm events, $N(\BpBm)=(197.2\pm3.1) \times 10^6$,
is determined using $N_{\BB}$ and the ratio of $\Gamma(\Y\to\BpBm)/\Gamma(\Y\to\BoBob)$ ($=1.065\pm0.026$)~\cite{PDG}.

Since the reconstruction efficiencies vary slightly for different resonances,
the average efficiency is calculated by weighing the
accepted and generated events by $S(x,y)$ with the values for the parameters of
our nominal Dalitz plot model (discussed below):
\begin{eqnarray}
\bar{\eff} = \frac{\sum_{i=1}^{\Nrec}S(x_i,y_i)\times w_i}{\sum_{j=1}^{\Ngen}S(x_j,y_j)},
\label{epsilon}
\end{eqnarray}
where $w_i$ is the correction factor which depends on $x$ and $y$ due to particle identification efficiency.
The value $\bar{\eff} = (8.72\pm0.05)$\% is obtained using this method.

The measured total branching fraction is
$\Br(\BDC) = (1.08\pm 0.03)\times 10^{-3}$,
where the stated error refers to the statistical uncertainty only.
A full discussion of the systematic uncertainties follows below.

\subsection{Dalitz plot analysis results}

The Dalitz plot distribution for data is shown in Fig. 10.
Since the composition of events in the Dalitz plot and their distributions are not 
known {\it a priori}, we have
tried a variety of different assumptions.  In particular,
we test the inclusion of various components,
such as the virtual \Dv and \Bv as well as S-, P- and D-wave modeling of the nonresonant component,
in addition to the expected components of $D^{*0}_2$, $D^{*0}_0$ and background.
The D-wave nonresonant term does not improve the goodness-of-fit and the fraction of D-wave nonresonant contribution
is close to 0.
The results of these tests with variations of the models are summarized in Table~\ref{tab:fitdataOmission}.
\begin{table*}
\begin{center}
\caption{Fit results for the masses, widths, fit fractions and phases from the Dalitz plot analysis
of $\BDC$ for different models. The errors are statistical only.
The magnitude and phase of the \Dtz amplitude are fixed to 1 and 0, respectively.
The background fraction is fixed to 30.4\% as described in Section~\ref{sec:EventSelection}.
The nominal fit corresponds to model 1.
The labels, S-NR and P-NR, denote the S-wave nonresonant and P-wave nonresonant contributions, respectively.}
\begin{tabular}{|l|c|c|c|c|c|c|c|}
\hline
Parameter                  & Model 1         & Model 2         & Model 3         & Model 4         & Model 5         & Model 6         & Model 7          \\
\hline
$m_{\Dtz}$ (\mevcc)        & $2460.4\pm1.2$  & $2460.2\pm1.0$  & $2459.1\pm1.0$  & $2460.1\pm1.1$  & $2461.5\pm1.2$  & $2458.1\pm1.1$  & $2457.4\pm1.0$   \\
$\Gamma_{\Dtz}$ (\mev)     & $41.8  \pm2.5$  & $41.7  \pm2.4$  & $41.1  \pm2.4$  & $41.8  \pm2.4$  & $42.0  \pm2.5$  & $41.8  \pm2.4$  & $41.7  \pm2.4$   \\
$m_{\Dzz}$ (\mevcc)        & $2297  \pm8  $  & $2309  \pm7  $  & $2297  \pm7 $   & $2312  \pm10 $  & $2307  \pm11 $  & $2270  \pm8  $  & $2273  \pm5  $   \\
$\Gamma_{\Dzz}$ (\mev)     & $273   \pm12 $  & $285   \pm11 $  & $288   \pm12 $  & $289   \pm20 $  & $313   \pm21 $  & $262   \pm12 $  & $276   \pm10 $   \\
$f_{\Dtz}$ (\%)            & $32.2  \pm1.3$  & $30.8  \pm1.2$  & $ 31.5 \pm1.1$  & $30.7  \pm1.5$  & $32.6  \pm1.3$  & $ 32.9 \pm1.3$  & $ 30.9 \pm1.1$   \\
$\phi_{\Dtz}$ (rad)        & 0.0 (fixed)     & 0.0 (fixed)     & 0.0 (fixed)     & 0.0 (fixed)     & 0.0 (fixed)     & 0.0 (fixed)     & 0.0 (fixed)      \\
$f_{\Dzz}$ (\%)            & $ 62.8 \pm2.5$  & $ 59.0 \pm2.1$  & $ 57.5 \pm1.7$  & $ 57.0 \pm4.5$  & $ 88.0 \pm8.1$  & $ 64.8 \pm2.2$  & $ 69.7 \pm1.1$   \\
$\phi_{\Dzz}$ (rad)        & $-2.07 \pm0.06$ & $-2.06 \pm0.05$ & $-2.01 \pm0.05$ & $-2.00 \pm0.12$ & $-2.14 \pm0.10$ & $-1.96 \pm0.06$ & $-2.00 \pm0.05$  \\
$f_{\Dv}$ (\%)             & $10.1 \pm1.4$   & $11.3 \pm1.5$   & $ 9.0  \pm1.2$  & $11.0 \pm1.5$   & $ 9.6 \pm1.3$   &                 &                  \\
$\phi_{\Dv}$ (rad)         & $ 3.00\pm0.12$  & $ 2.99\pm0.08$  & $ 3.17 \pm0.10$ & $ 3.05\pm0.12$  & $ 2.82\pm0.17$  &                 &                  \\
$f_{\Bv}$ (\%)             & $ 4.6 \pm2.6$   & $ 1.4 \pm0.5$   &                 & $ 1.7 \pm0.8$   & $12.2 \pm5.4$   & $2.2  \pm1.4$   &                  \\
$\phi_{\Bv}$ (rad)         & $ 2.80\pm0.21$  & $ -2.43\pm0.28$ &                 & $ -2.33\pm0.28$ & $ 2.52\pm0.25$  & $2.28\pm0.38$   &                  \\
$f_{\mbox{P-NR}}$ (\%)     & $ 5.4\pm2.4$    &                 & $1.6  \pm0.4$   &                 & $12.6\pm4.0$    & $12.7\pm3.1$    &                  \\
$\phi_{\mbox{P-NR}}$ (rad) & $-0.89\pm0.18$  &                 & $-1.46\pm0.20$  &                 & $-0.84\pm0.12$  & $-0.71\pm0.10$  &                  \\
$f_{\mbox{S-NR}}$ (\%)     &                 &                 &                 & $0.3\pm0.3$     & $5.2\pm3.8$     &                 &                  \\
$\phi_{\mbox{S-NR}}$ (rad) &                 &                 &                 & $-0.77\pm0.49$  & $3.30\pm0.23$   &                 &                  \\
$f_{bg}$ (\%)              & $30.4$(fixed)   & $30.4$(fixed)   &  $30.4$(fixed)  & $30.4$(fixed)   & $30.4$(fixed)   & $30.4$(fixed)   & $30.4$(fixed)    \\
\hline
NLL                        & 22970           & 22982           &  22977          & 22982           & 22964           &  23046          & 23125            \\
$\chi^2/\mbox{NDF}$        & $220/153$       & $240/152$       &  $236/154$      & $239/153$       & $216/150$       & $328/160$       & $454/161$        \\
\hline
\end{tabular}
\label{tab:fitdataOmission}
\end{center}
\end{table*}
Of these models, model 1 produces the best fit quality with the smallest number of components,
and we choose it as the nominal fit model.
The components considered in this fit model are 
$\Dtz$, $\Dzz$, $\Dv$, $\Bv$ and P-wave nonresonant.
The P-wave nonresonant component is an addition to the fit model used in the previous measurement from Belle~\cite{belle-prd}.
The sum of the fractions ($115\pm5$)\% for the nominal fit
differs from 100\% because of destructive interferences among the amplitudes.
The $\chi^2/\mbox{NDF}$ for the nominal fit is $220/153$.
To better understand the large $\chi^2/\mbox{NDF}$, we look at the contributions to
the total $\chi^2$ from individual cells. We find four cells with $\chi^2>7$, which inflate the
total $\chi^2$. The central points in these cells are at: ($6.83,-0.722$), ($6.83,-0.611$), ($6.83,0.5$)
and ($8.08,-0.722$), where the first value is \MMminDP, and the second is \ct.
In order to determine the effect on the fitted parameters from these cells, we repeat the nominal
fit with these cells excluded.
The resulting $\chi^2/\mbox{NDF}$ is $182/149$, corresponding to a probability of 3.4\%.
Assuming these large $\chi^2$ contributions are caused by an unknown
systematic problem, removing them from the fit is reasonable.
However, under the assumption that these high $\chi^2$ contributions
have a statistical origin, the $\chi^2$ probability is 0.04\%~\cite{prob}.
The low probability indicates that a model more complex than the
isobar model may be necessary to describe the characteristics of the data.
The differences in the fitted $D^*_2$ and $D^*_0$ parameters, when these cells are
included or excluded, are assigned to systematic uncertainties,
and are much smaller than the statistical uncertainties.
The removal of these cells does not affect the choice of model 1
as the nominal fit from Table~\ref{tab:fitdataOmission}.

Ref.~\cite{dpi} argues for an addition of a $D\pi$ S-wave state near the
$D\pi$ system threshold to the model of the $D\pi\pi$ final state.
We have performed tests using the models 1-4
in Table~\ref{tab:fitdataOmission} with the $D^*_v$ replaced by a $D\pi$ S-wave state.
Two different parametrizations for $D\pi$ S-wave state amplitude are used:
one is the function given by Eq.(8) of Ref.~\cite{dpi} with
the numerator set to constant,
the other function is the relativistic Breit-Wigner given by Eq. (17).
Among the tests we have performed with these parameterizations,
the model with $D^*_2$, $D^*_0$, $D\pi$
S-wave (using Eq.(8) of Ref.~\cite{dpi}), $B^*_v$ and
P-wave nonresonant gives the best fit with NLL and $\chi^2/\mbox{NDF}$ values
of 22997 and $271/151$, respectively, which are worse than those of the nominal fit
even when allowing the $D\pi$ S-wave's parameters to vary.
Each of these models also requires large fractions of $D^*_0$.

\begin{figure}[b]
\centering
\epsfig{width=0.44\textwidth,file=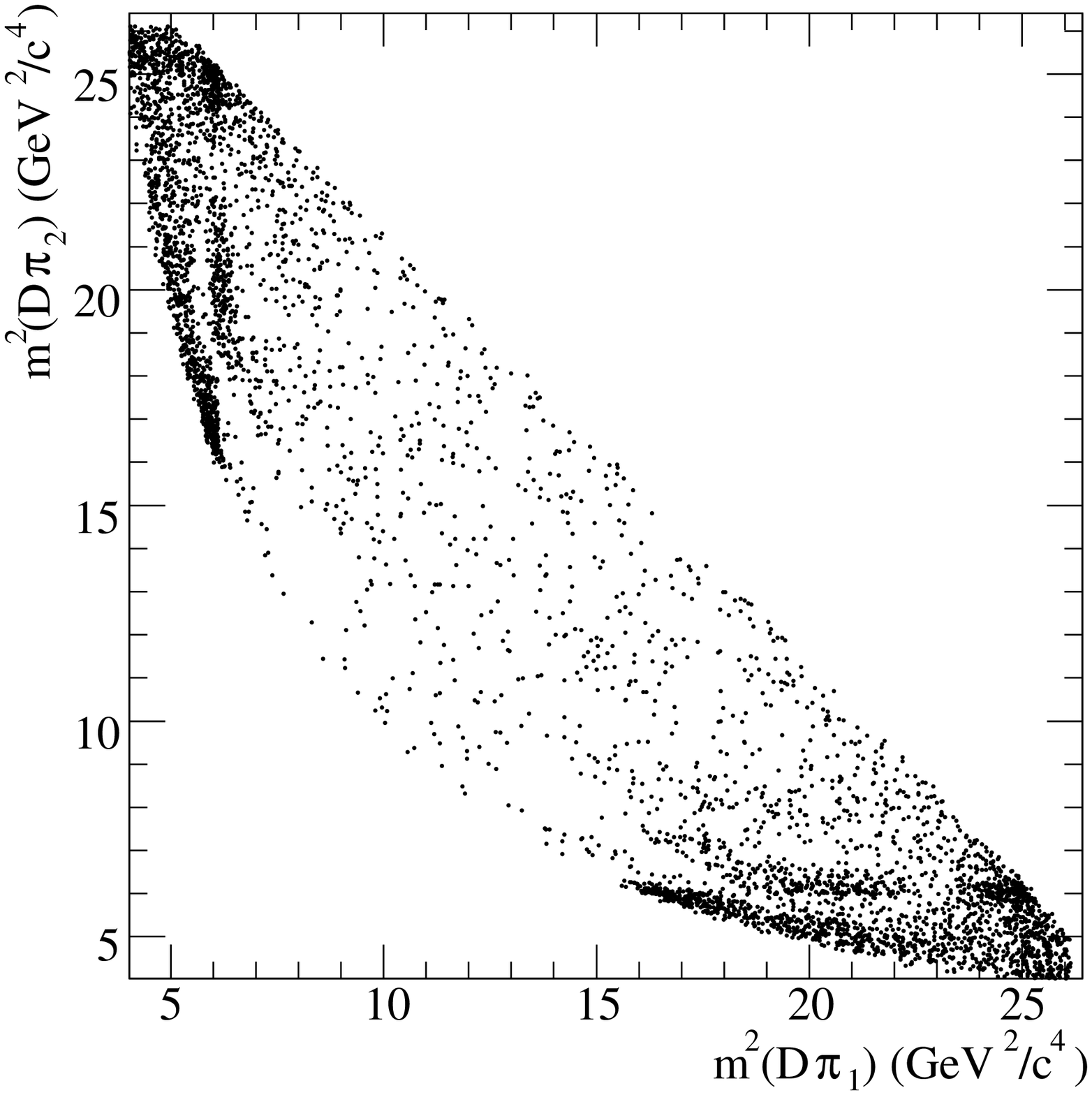}
\label{fig:datadalitz}
\caption{Data Dalitz Plot for \BDC.}
\end{figure}
\begin{figure}[t]
\centering
\epsfig{width=0.477\textwidth,file=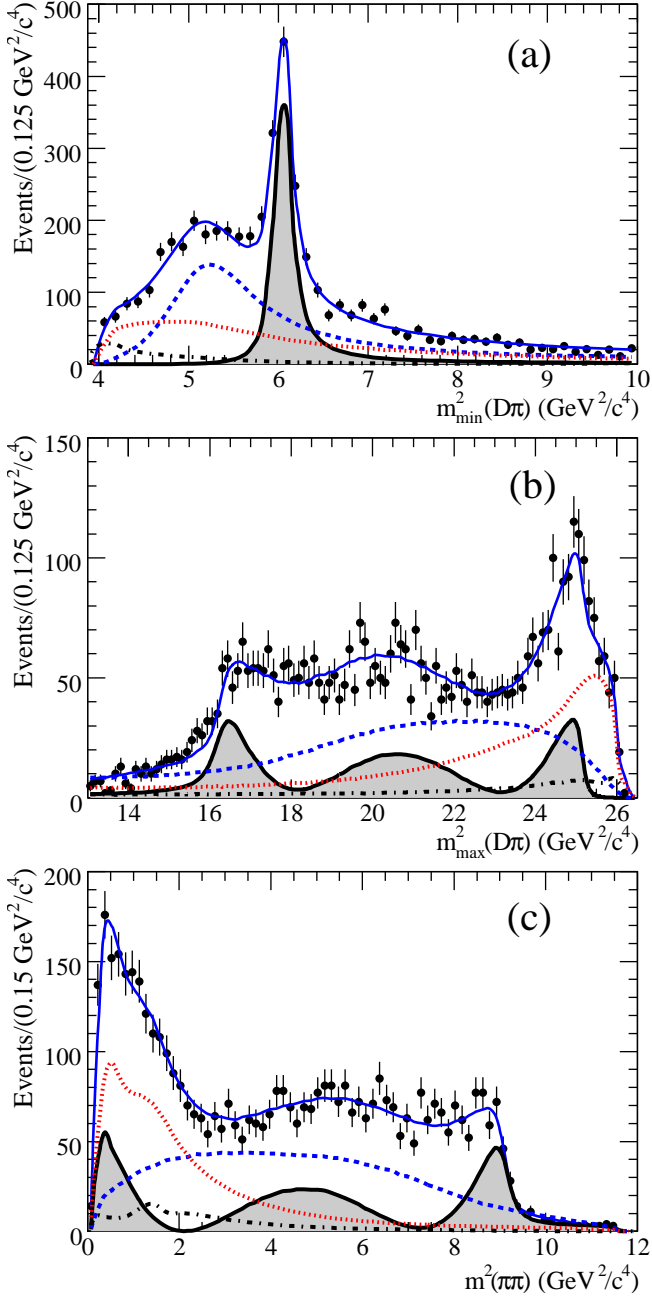}
\label{fig:datafit1}
\caption{Result of the nominal fit to the data: projections on 
(a) \MMminDP, (b) \MMmaxDP  and (c) \MMPP.
The points with error bars are data, the solid curves represent the nominal fit.
The shaded areas show the \Dtz contribution,
the dashed curves show the \Dzz signal, the dash-dotted curves show the \Dv and \Bv signals,
and the dotted curves show the background.}
\end{figure}
\begin{figure}[t]
\centering
\epsfig{width=0.477\textwidth,file=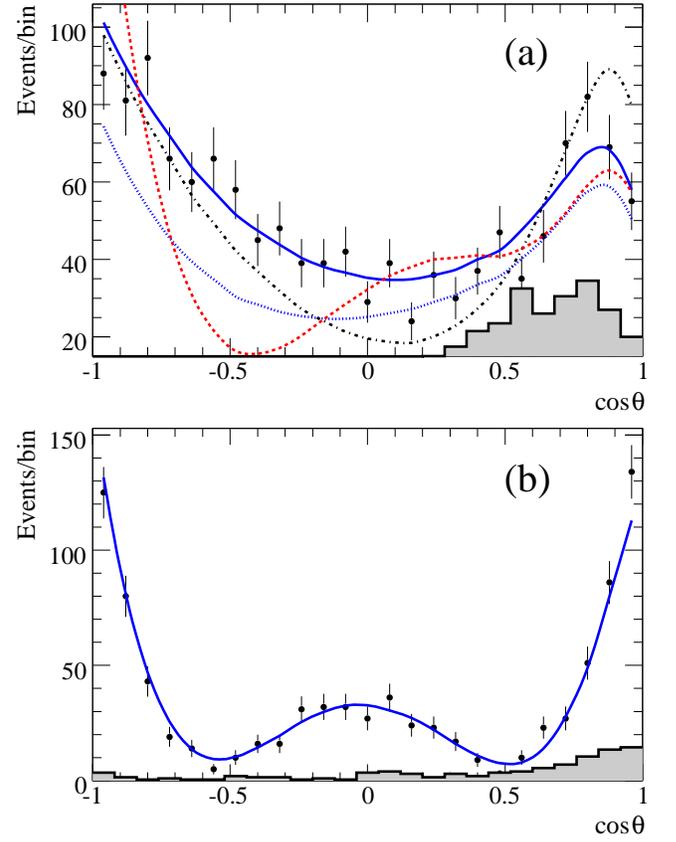}
\label{fig:datafit2}
\caption{Result of the nominal fit to the data: the \ct distributions for 
(a) $4.5 < m^2(D\pi) < 5.5 \gevccs$ region and 
(b) $5.9 < m^2(D\pi) < 6.2 \gevccs$ region.
The points with error bars are data, the solid curves represent the nominal fit.
The dashed, dash-dotted and dotted curves in (a) show the fit of hypotheses 2-4 in Table~\ref{d0test}, respectively.
The shaded histograms
show the \ct distributions from \de sidebands in data.}
\end{figure}

The nominal fit model results in the following branching fractions: 
$\Br(B^- \to D^{*0}_2 \pi^-) \times \Br(D^{*0}_2 \to D^+\pi^-) = (3.5\pm0.2) \times 10^{-4}$ and
$\Br(B^- \to D^{*0}_0 \pi^-) \times \Br(D^{*0}_0 \to D^+\pi^-) = (6.8\pm0.3) \times 10^{-4}$,
where the errors are statistical only.
A full discussion of the systematic uncertainties follows below.

Fig.~11a, 11b and 11c show the 
\MMminDP, \MMmaxDP and \MMPP projections respectively,
while Fig.~12a and 12b show the
\ct distributions for the $D^{*0}_0$ and $D^{*0}_2$ mass regions, respectively.
The distributions in Fig.~11 and 12 show good agreement between the data and the fit.
The angular distribution in the $D^{*0}_2$ mass region
is clearly visible and is consistent with the expected D-wave distribution of $\mid \cos^2\theta-1/3 \mid ^2$ for a spin-2 state.
In addition, the $D^{*0}_0$ signal and the reflection of $D^{*0}_2$
can be easily distinguished in the \MMminDP and \MMmaxDP projection, respectively.
The lower edge of  \MMminDP is better described with \Dv component included than without.

Table~\ref{d0test} shows the NLL and $\chi^2/\mbox{NDF}$ values for the nominal fit and 
for the fits with the broad resonance \Dzz excluded or with the
$J^P$ of the broad resonance replaced by other quantum numbers.
In all cases, the NLL and $\chi^2/\mbox{NDF}$ values are significantly worse than that of the nominal fit.
Fig.~12a illustrates the helicity distributions in the $D^{*0}_0$ mass region
from hypotheses 2-4; clearly the nominal fit gives the best description of the data.
We conclude that a broad spin-0 state \Dzz is required in the fit to the data.
The same conclusion is obtained when performing the 
same test on Models 2-5.

\begin{table}[htbp]
\begin{center}
\caption{Comparison of the models with different resonances composition.
The labels, S-NR and P-NR, denote the S-wave nonresonant and P-wave nonresonant contributions, respectively.}
\begin{tabular}{|c|l|c|c|}
\hline
Hypothesis & Model   & NLL & $\chi^2/\mbox{NDF}$\\
\hline
~ & Model 1 (nominal fit)             &  22970 & $220/153 $ \\
\hline
1 & $\Dtz, \Dv, \Bv$, P-NR            &  23761 & $1171/143$ \\
2 & $\Dtz, \Dv, \Bv$, P-NR, $(2^+) $  &  23699 & $ 991/144$ \\
3 & $\Dtz, \Dv, \Bv$, P-NR, $(1^-) $  &  23427 & $ 638/135$ \\
4 & $\Dtz, \Dv, \Bv$, P-NR, S-NR      &  23339 & $ 652/157$ \\
\hline
\end{tabular}
\label{d0test}
\end{center}
\end{table}

\section{SYSTEMATIC UNCERTAINTIES}
\label{sec:Systematics}

\subsection{Uncertainties on $\Br(\Bm \to D^{+}\pim\pim)$}

As listed in Table~IV, the systematic 
error on the measurement of the total \BDC branching fraction is due to the uncertainties on the following quantities:
the number of \BpBm events in the initial sample, the charged track reconstruction and identification efficiencies,
and the $\Dp \to \KPP$ branching fraction. The uncertainty in the \de background shape, the uncertainty in the average efficiency
due to the fit models and 
a possible fit bias also contribute to the systematic error.
\begin{table}[bht]
\begin{center}
\caption{Summary of systematic uncertainties (relative errors in \%) in the measurement of the total branching fraction.}
\begin{tabular}{|l|c|}
\hline
Systematic Source        & $\frac{\Delta\Br(\BDC)}{\Br(\BDC)}$ (\%)\\
\hline
Number of \BpBm events     & $1.6$\\
Tracking efficiencies    & $2.5$ \\
PID                      & $1.5$\\
\de background shape     & $1.3$ \\
$D^+$ branching fraction & $2.3$ \\
Fit models               & $0.7$ \\
Fit bias                 & $1.0$ \\
\hline
Total Systematics        & $4.4$\\
\hline
\end{tabular}
\end{center}
\label{table:brsys}
\end{table}

The uncertainty on the number of \BpBm events is determined using the uncertainties on
$\Gamma(\Y\to\BpBm)/\Gamma(\Y\to\BoBob)$~\cite{PDG} and integrated luminosity (1.1\%).
The uncertainty on the input \Dp branching fraction is taken
from~\cite{cleodpp}.
The uncertainty in the \de background shape
is estimated by comparing the signal yields between fitting the \de distribution
with a linear background shape and with higher-order (second and third-order) polynomials.
The uncertainty in the fit models is estimated by comparing the average efficiencies in Eq.~(\ref{epsilon}) using
Models 2-5 of Table~\ref{tab:fitdataOmission}.
The fit bias is estimated to be less than 1\% by comparing
the generated and the fitted value of $\Br(\Bm \to D^{+}\pim\pim)$ from resonant and continuum MC samples.

\subsection{Uncertainties on Dalitz plot analysis results}

The sources of systematic uncertainties that affect the results of the Dalitz Plot analysis
are summarized in Table~V.
These uncertainties are added in quadrature, as they are uncorrelated, to obtain the total systematic error.

The uncertainties due to the background
parameterization are estimated by comparing the results from the nominal
fit with those obtained when the background shape parameters are allowed
to float in the fit.
The errors from the uncertainty in the background fraction are estimated by comparing
the fit results when the background fraction is changed by its statistical error.
We vary the set of cuts on \de, \mes, $R_2$,
\Cth and mass of \Dp, which increase the number of signal events by 25\%
and the background fraction to 36.5\%, and repeat the fits.
The difference in the fit results is taken as an estimate of the systematic uncertainty due to the event selection.
Fit biases are studied using 1248 parameterized MC samples and
262 MC samples with full event reconstruction.
Small biases are observed for some of the parameters.
We combine these biases with those coming from high $\chi^2$
cells, as discussed in the previous section, in quadrature to obtain the
total systematic contribution from the fit bias.
The uncertainties in PID are obtained by comparing the nominal fit results with those obtained when
the PID corrections to the reconstruction efficiency are varied according to their uncertainties.
The uncertainties in the efficiency and signal resolution parametrization 
are found to be negligible using the fits to the reconstructed MC samples.

In addition to the above systematic uncertainties, we also estimate a 
model-dependent uncertainty that comes from the uncertainty in the composition of the signal model and
the uncertainty in the Blatt-Weisskopf barrier factors.
The model-dependent uncertainties are estimated by comparing the fit results with
Models 2-5 in Table~\ref{tab:fitdataOmission}
and by varying the radius of the barrier, $r^\prime$ and $r$ in Eqs.~(\ref{barrier1})-(\ref{barrier3})
from 0 to 5~$\mathrm{(GeV/c)}^{-1}$.

\begin{table*}[htb]
\begin{center}
\caption{Summary of systematic uncertainties in the masses, widths and fit fractions of the $\Dtz$ and $\Dzz$ and the phase of $\Dzz$.}
\begin{tabular}{|l|c|c|c|c|c|c|c|}
\hline
Systematic Source           & $\Delta m_{\Dtz}$ & $\Delta \Gamma_{\Dtz}$ & $\Delta m_{\Dzz}$ & $\Delta \Gamma_{\Dzz}$ & $\Delta f_{\Dtz}$ & $\Delta f_{\Dzz}$ & $\Delta \phi_{\Dzz}$\\
~                           & (\mevcc)   & (\mev)          & (\mevcc)   & (\mev) & (\%) & (\%) & (rad) \\
\hline
Background parameterization & 1.0 & 1.1 &  3 &  5 & 1.2 &  0.0 & 0.04 \\
Background fraction         & 0.1 & 0.4 &  2 &  1 & 0.4 &  0.4 & 0.00 \\
Event selection             & 0.6 & 1.6 &  1 & 14 & 0.3 &  0.8 & 0.08 \\
Fit bias                    & 0.3 & 0.7 &  4 &  8 & 0.7 &  1.4 & 0.02 \\
PID efficiency              & 0.0 & 0.1 &  0 &  0 & 0.0 &  0.1 & 0.01 \\
\hline
Total systematic error      & 1.2 & 2.1 &  5 & 17 & 1.5 &  1.7 & 0.09 \\
\hline
Fit models                 & 1.3 & 0.7 & 15 & 40 & 1.5 & 17.2 & 0.07 \\
$r$ constant                & 1.4 & 1.9 & 12 & 21 & 3.8 &  7.8 & 0.17 \\
\hline
Total model-dependent error & 1.9 & 2.0 & 19 & 45 & 4.1 & 18.9 & 0.18 \\
\hline
\end{tabular}
\end{center}
\label{table:masswidthsystematics}
\end{table*}

\section{SUMMARY}
\label{sec:Summary}

In conclusion, we measure
the total branching fraction of the \BDC decay to be
\begin{eqnarray}
\Br(\BDC) = (1.08 \pm 0.03 \pm 0.05) \times 10^{-3}, \nonumber
\end{eqnarray}
where the first error is statistical and the second is
systematic.

Analysis of the \BDC Dalitz plot using the isobar model confirms the existence
of a narrow \Dtz and a broad \Dzz resonance as predicted by Heavy Quark Effective Theory.
The mass and width of \Dtz are determined to be:
\begin{eqnarray}
m_{\Dtz} &=& (2460.4\pm1.2\pm1.2\pm1.9) \mevcc  \mbox{~and} \nonumber\\
\Gamma_{\Dtz} &=& (41.8\pm2.5\pm2.1\pm2.0) \mev, \nonumber
\end{eqnarray}
respectively, while for the \Dzz they are:
\begin{eqnarray}
m_{\Dzz} &=& (2297\pm8\pm5\pm19) \mevcc \mbox{~and} \nonumber\\
\Gamma_{\Dzz} &=& (273\pm12\pm17\pm45) \mev, \nonumber
\end{eqnarray}
where the first and second errors reflect the statistical
and systematic uncertainties, respectively, the third one
is the uncertainty related to the assumed composition of signal events
and the Blatt-Weisskopf barrier factors.
The measured masses and widths of both states are consistent
with the world averages~\cite{PDG} and the predictions of some theoretical models~\cite{isgur,ebert,kalashnikova}.

We have also obtained exclusive branching fractions for $D^{*0}_2$ and $D^{*0}_0$ production:
\begin{eqnarray}
\Br(B^- &\to& D^{*0}_2 \pi^-) \times \Br(D^{*0}_2 \to D^+\pi^-) \nonumber\\
&=& (3.5\pm0.2\pm0.2\pm0.4) \times 10^{-4} \mbox{~and} \nonumber\\
\Br(B^- &\to& D^{*0}_0 \pi^-) \times \Br(D^{*0}_0 \to D^+\pi^-) \nonumber\\
&=& (6.8\pm0.3\pm0.4\pm2.0) \times 10^{-4}. \nonumber
\end{eqnarray}

Our results for the masses, widths and branching fractions are consistent
with but more precise than previous measurements performed by Belle~\cite{belle-prd}.

The relative phase of the scalar and tensor amplitude is measured to be
\begin{eqnarray}
\phi_{D^{*0}_0} = -2.07\pm0.06\pm0.09\pm0.18 \mbox{~rad}. \nonumber
\end{eqnarray}

\begin{acknowledgments}
We are grateful for the 
extraordinary contributions of our \peptwo\ colleagues in
achieving the excellent luminosity and machine conditions
that have made this work possible.
The success of this project also relies critically on the 
expertise and dedication of the computing organizations that 
support \babar.
The collaborating institutions wish to thank 
SLAC for its support and the kind hospitality extended to them. 
This work is supported by the
US Department of Energy
and National Science Foundation, the
Natural Sciences and Engineering Research Council (Canada),
the Commissariat \`a l'Energie Atomique and
Institut National de Physique Nucl\'eaire et de Physique des Particules
(France), the
Bundesministerium f\"ur Bildung und Forschung and
Deutsche Forschungsgemeinschaft
(Germany), the
Istituto Nazionale di Fisica Nucleare (Italy),
the Foundation for Fundamental Research on Matter (The Netherlands),
the Research Council of Norway, the
Ministry of Education and Science of the Russian Federation, 
Ministerio de Educaci\'on y Ciencia (Spain), and the
Science and Technology Facilities Council (United Kingdom).
Individuals have received support from 
the Marie-Curie IEF program (European Union) and
the A. P. Sloan Foundation.
\end{acknowledgments}


\begin{thebibliography}{99}
\bibitem{conjugate} Charged conjugate states are implied throughout the paper.
\bibitem{prob} 0.04\% is the probability to obtain a $\chi^2$ of 182 or greater
               for a collection of 153 standard normal-distributed numbers,
               of which the four largest positive deviations have been removed.
\end{thebibliography}
\end{document}